\documentclass[12pt]{article}

\usepackage{epsfig}

\usepackage{graphicx}
\usepackage{amsfonts}
\usepackage{amssymb}
\usepackage{float}

\def\gtwid{\mathrel{\raise.3ex\hbox{$>$\kern-.75em\lower1ex\hbox{$\sim$}}}}
\def\ltwid{\mathrel{\raise.3ex\hbox{$<$\kern-.75em\lower1ex\hbox{$\sim$}}}}
\def\square{\kern1pt\vbox{\hrule height 1.2pt\hbox{\vrule width 1.2pt\hskip 3pt
   \vbox{\vskip 6pt}\hskip 3pt\vrule width 0.6pt}\hrule height 0.6pt}\kern1pt}

\begin{document}

\begin{titlepage}

\begin{flushright}
UFIFT-QG-17-05, CCTP-2017-10
\end{flushright}

\vskip 1cm

\begin{center}
{\bf From Non-trivial Geometries to Power Spectra and Vice Versa}
\end{center}

\vskip 1cm

\begin{center}
D. J. Brooker$^{1*}$, N. C. Tsamis$^{2\star}$ and R. P. Woodard$^{1\dagger}$
\end{center}

\begin{center}
\it{$^{1}$ Department of Physics, University of Florida,\\
Gainesville, FL 32611, UNITED STATES}
\end{center}

\begin{center}
\it{$^{2}$ Institute of Theoretical Physics \& Computational Physics, \\
Department of Physics, University of Crete, \\
GR-710 03 Heraklion, HELLAS}
\end{center}

\vspace{1cm}

\begin{center}
ABSTRACT
\end{center}
We review a recent formalism which derives the functional forms 
of the primordial -- tensor and scalar -- power spectra of scalar 
potential inflationary models. The formalism incorporates the case
of geometries with non-constant first slow-roll parameter. Analytic
expressions for the power spectra are given that explicitly display
the dependence on the geometric properties of the background. 
Moreover, we present the full algorithm for using our formalism,
to reconstruct the model from the observed power spectra. Our 
techniques are applied to models possessing ``features" in their
potential with excellent agreement.

\begin{flushleft}
PACS numbers: 04.50.Kd, 95.35.+d, 98.62.-g
\end{flushleft}

\vskip .5cm

\begin{flushleft}
$^{*}$ e-mail: djbrooker@ufl.edu \\
$^{\star}$ e-mail: tsamis@physics.uoc.gr \\
$^{\dagger}$ e-mail: woodard@phys.ufl.edu
\end{flushleft}

\end{titlepage}

\section{Introduction}

We shall assume that inflation is described by general
relativity minimally coupled to a scalar field $\varphi(x)$
with a self-interacting potential $V(\varphi)$:
\footnote{Hellenic indices take on spacetime values 
while Latin indices take on space values. Our metric
tensor $g_{\mu\nu}$ has spacelike signature
$( - \, + \, + \, +)$ and our curvature tensor equals
$R^{\alpha}_{~ \beta \mu \nu} \equiv 
\Gamma^{\alpha}_{~ \nu \beta , \mu} +
\Gamma^{\alpha}_{~ \mu \rho} \, 
\Gamma^{\rho}_{~ \nu \beta} -
(\mu \leftrightarrow \nu)$.}
\begin{equation}
\mathcal{L} \, = \,
\frac{1 \sqrt{-g}}{16 \pi G} R \sqrt{-g} 
- \frac12 \partial_{\mu} \varphi \, 
\partial_{\nu} \varphi \, g^{\mu\nu} \sqrt{-g} 
- V(\varphi) \, \sqrt{-g} 
\;\; . \label{Lagrangian}
\end{equation}
The theory described by (\ref{Lagrangian}) predicts 
the generation of tensor \cite{Starobinsky:1979ty}
and scalar \cite{Mukhanov:1981xt}
perturbations. These predictions provide the main test
for the validity of such models \cite{Mukhanov:1990me,
Liddle:1993fq} as well as the reconstruction of the potential 
$V(\varphi)$ \cite{Lidsey:1995np}.
The class of spacetimes under consideration is characterized
by the scale factor $a(t)$ and, hence, the Hubble parameter
$H(t)$ and the first slow-roll parameter $\epsilon(t)$:
\begin{equation}
ds^2 = -dt^2 + a^2(t) \, d\vec{x} \!\cdot\! d\vec{x} 
\quad \Longrightarrow \quad
H(t) \equiv \frac{\dot{a}}{a} 
\quad , \quad 
\epsilon(t) \equiv -\frac{\dot{H}}{H^2} 
\;\; . \label{geometry}
\end{equation}

We shall study the tree order tensor and scalar primordial
power spectra, $\Delta^2_{h}(k)$ and $\Delta^2_{\mathcal{R}}(k)$ 
respectively. They are known in terms of the constant amplitudes 
approached by their mode functions, $u(t,k)$ and $v(t,k)$
respectively, after the first horizon crossing time $t_k$
\cite{Woodard:2014jba}:
\begin{eqnarray}
\Delta^2_{h}(k) &\!\! = \!\!& 
\frac{k^3}{2\pi^2} \times 32\pi G \times 2 \times 
\Big\vert u(t,k)\Big\vert^2_{t \gg t_k} 
\;\; , \label{Dh} \\
\Delta^2_{\mathcal{R}}(k) &\!\! = \!\!& 
\frac{k^3}{2 \pi^2} \times 4 \pi G \times 
\Big\vert v(t,k)\Big\vert^2_{t \gg t_k} 
\;\; , \label{DR}
\end{eqnarray}
where $k = H(t_k) a(t_k)$. The time evolution equations 
obeyed by these mode functions:
\begin{eqnarray}
\ddot{u} + 3 H \dot{u} + \frac{k^2}{a^2} u = 0 
& , &
u \dot{u}^* \!- \dot{u} u^* = \frac{i}{a^3} 
\;\; , \label{ueqns} \\ 
\ddot{v} + \Bigl( 3 H + \frac{\dot{\epsilon}}{\epsilon} \Bigr) 
\dot{v} + \frac{k^2}{a^2} v = 0 
& , & 
v \dot{v}^* \!- \dot{v} v^* = \frac{i}{\epsilon a^3} 
\;\; , \label{veqns}
\end{eqnarray}
cannot be solved exactly and we must resort to complicated
numerical techniques for realistic inflationary models.

It is evident from equations (\ref{ueqns}-\ref{veqns}) that
constant solutions exist when $\frac{k^2}{a^2}$ becomes
negligible. Exact solutions are known for $\epsilon(t) 
= \epsilon_0$:
\begin{eqnarray}
u_0(t,k;\epsilon_0) &\!\! = \!\!& 
\sqrt{\frac{\pi}{4 k a^2(t)} \, z(t)} \; 
H^{(1)}_{\nu}\bigl( z(t) \bigr)
\quad , \quad
v_0(t,k;\epsilon_0) \, = \,
\frac{u_0(t,k;\epsilon_0)}{\sqrt{\epsilon_0}} 
\;\; , \nonumber \\
z(t) &\!\!\equiv \!\!& 
\frac{k}{(1 - \epsilon_0) H(t) \, a(t)}
\quad , \quad 
\nu \, \equiv \, 
\frac12 + \frac{1}{1 - \epsilon_0} 
\;\; . \label{consteps}
\end{eqnarray}
However, no constant value of $\epsilon(t)$ seems to be 
consistent with the data, cf. Figure 12 of \cite{Ade:2015lrj}.
Achieving more realism involves consideration of geometries 
with non-constant $\epsilon(t)$ \cite{Starobinsky:1992ts,
Wang:1997cw}. Therefore, we must go beyond the leading 
slow-roll approximation -- $\epsilon(t) = \epsilon_0 \ll 1 
\; , \epsilon'(t) = 0$ -- to the power spectra:
\begin{equation}
\Delta^2_{h}(k) \Big\vert_{\rm leading} =  
\frac{16 G H^2(t_k)}{\pi} 
\quad , \quad
\Delta^2_{\mathcal{R}}(k) \Big\vert_{\rm leading} = 
\frac{G H^2(t_k)}{\pi \epsilon(t_k)} 
\;\; . \label{Dleading}
\end{equation}
which, although qualitatively accurate over most of the observed 
spectrum, does not provide a good description of features, for 
example, the power deficit at $\ell = 22$ and the excess at 
$\ell = 40$ visible in Figure 1 \cite{Ade:2015xua}.  
\footnote{It should be noted that these are $3\sigma$
deviations and more accuracy is needed to establish them
as true physical results.}
\begin{figure}[H]
\includegraphics[width=13cm,height=7cm]{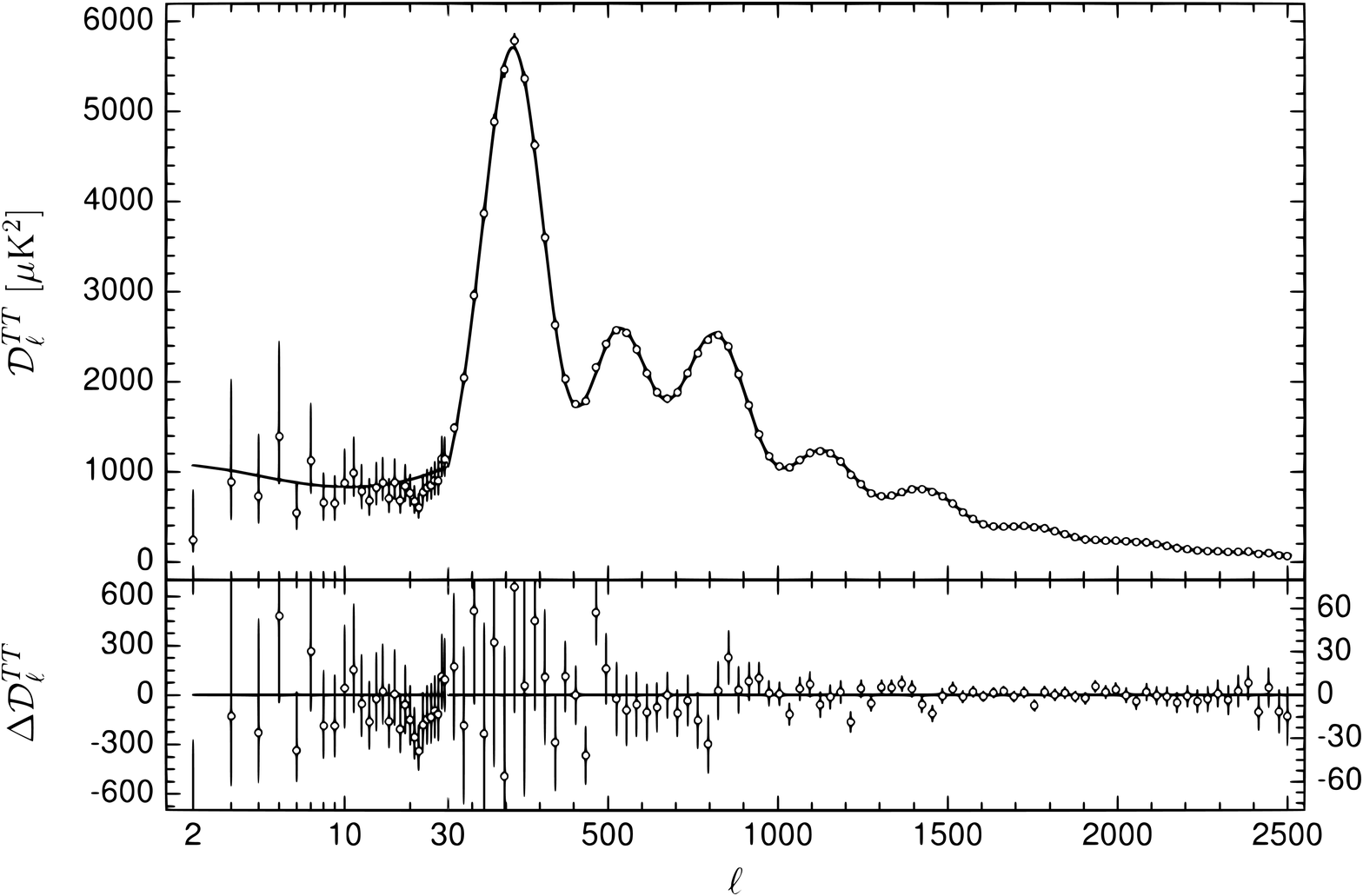}
\vspace{-0.2cm}
\caption{\footnotesize The PLANCK 2015 \cite{Ade:2015xua}
strength of temperature variations (vertical) against their 
angular sizes (horizontal). The line is the standard cosmological 
model, the dots are the data.}
\label{planck2015}
\end{figure}
The deviations do not disappear when we consider the local slow 
roll approximation -- $\epsilon(t) = \epsilon_0 \; , \epsilon'(t) 
= 0$ -- to the power spectra:
\begin{equation}
\Delta^2_{h}(k) \Big\vert_{\rm local} =  
\frac{16 G H^2(t_k)}{\pi} \times C[\epsilon_k] 
\quad , \quad
\Delta^2_{\mathcal{R}}(k) \Big\vert_{\rm local} = 
\frac{G H^2(t_k)}{\pi \epsilon(t_k)} \times C[\epsilon_k]
\;\; , \label{Dlocal}
\end{equation}
where the local slow-roll correction factor $C[\epsilon_0]$ is:
\begin{equation}
C[\epsilon_0] \, = \,
\frac{1}{\pi} \, 
\Gamma^2 \Big( \frac12 \!+\! \frac1{1 \!-\! \epsilon_0} \Big) 
\Bigl[ 2 (1 \!-\! \epsilon_0) \Bigr]^{\frac{2}{1-\epsilon_0}} 
\, \approx \, 1 - \epsilon_0
\;\; , \label{C}
\end{equation}
and its graph can be seen in Figure~\ref{C(e)}.
\begin{figure}[H]
\includegraphics[width=6.0cm,height=4.8cm]{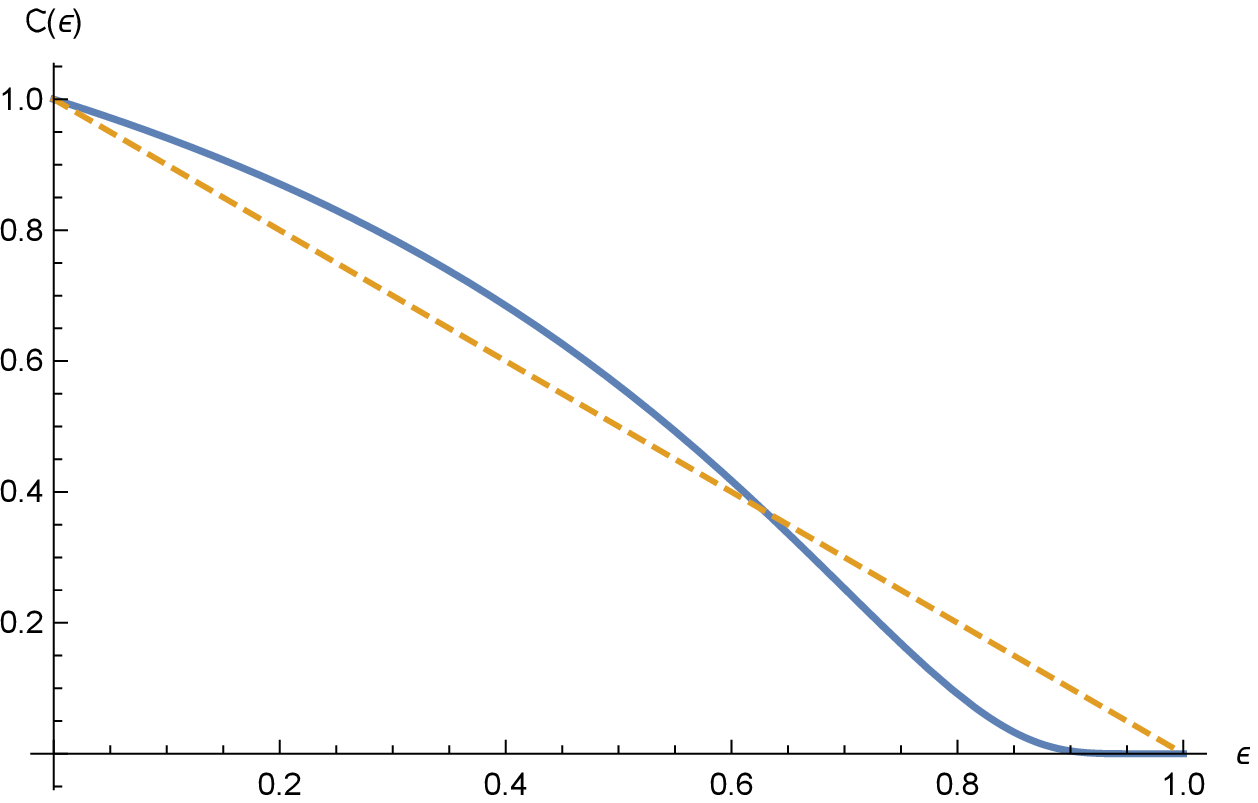}
\hspace{1cm}
\includegraphics[width=6.0cm,height=4.8cm]{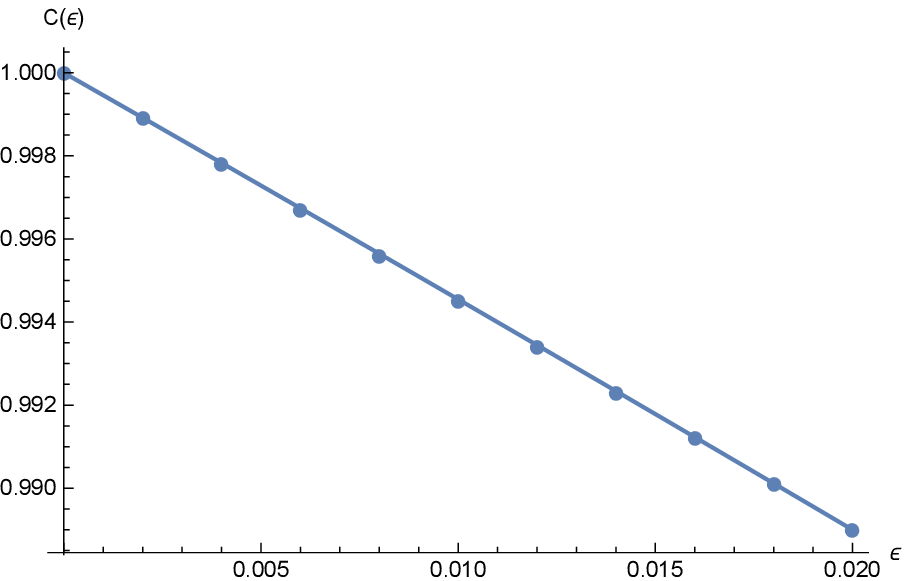}
\caption{\footnotesize The left hand graph shows the local 
slow-roll {\it constant} $\epsilon$ correction factor 
$C(\epsilon)$ {\it (solid blue)}, which was defined in 
expression (\ref{C}). Also shown is its global 
approximation of $\, 1 - \epsilon \,$ {\it (dashed yellow)} 
over the full inflationary range of $\, 0 \leq \epsilon < 1$. 
The right hand graph shows $C(\epsilon)$ {\it (solid blue)} 
versus the better approximation of $\, 1 - 0.55 \epsilon$ 
{\it (large dots)} relevant to the range 
$\, 0 \leq \epsilon < 0.02 \,$ favoured by current data.}
\label{C(e)}
\end{figure}

That said, we set some goals \cite{Brooker:2017kjd} when trying 
to develop a formalism that incorporates more generic inflationary 
models and goes beyond (\ref{Dleading}, \ref{Dlocal}) in a 
non-trivial and potentially quite interesting way:
\footnote{We are not concerned in this study with quantum 
corrections to the power spectra as their loop counting 
parameter is very small $(GH^2 << 10^{-11})$ and it is not 
clear what operators represent them at loop order \cite{Miao:2012xc}.}
\\ [3pt]
$\bullet \;$ The class of geometries studied must incorporate 
those with a varying $\epsilon(t)$. In particular, we should 
correctly describe transient effects in the power spectra due 
to the non-constancy of $\epsilon(t)$; these will eventually 
reside in the non-local dependence of the power spectrum 
on $\epsilon(t)$.
\\ [3pt]
$\bullet \;$ The formalism must be bi-directional: \\
(i) from spacetime 
$\{ a(t) \Rightarrow H(t) \Rightarrow \epsilon(t) \}$ 
to spectrum 
$\{ \Delta^2_{\mathcal{R}}(k) , \Delta^2_h(k) \}$ and, \\ 
(ii) from spectrum 
$\{ \Delta^2_{\mathcal{R}}(k) , \Delta^2_h(k) \}$
to spacetime 
$\{ a(t) \Rightarrow H(t) \Rightarrow \epsilon(t) \}$.
\\ [3pt]
$\bullet \;$ The formalism must be such that if numerical
methods need to be employed, they will be as efficient as
possible.

\section{From the Geometry to the CMBR}

In this Section we describe the steps that lead to 
the generalized expressions for the tensor and scalar
primordial power spectra. Because these steps are very 
similar in both cases, we shall be more detailed for 
the tensor case and rather compact for the scalar case.
\\ [7pt]
{\bf * The Tensor Spectrum} 
\\ [3pt]
{\it - Step 1: The optimal evolution variables.} 
\\ [2pt]
Elementary inspection of (\ref{Dh}) shows that the relevant 
quantity is not $u(t, k)$ evolving via (\ref{ueqns}) but
$M(t, k) \equiv \vert u(t, k) \vert^2$ evolving according
to \cite{Romania:2012tb,Brooker:2015iya}:
\begin{equation}
\frac{M''}{M} + (3 - \epsilon) \frac{M'}{M}
+ \frac{2 k^2}{a^2 H^2} 
- \frac12 \left( \frac{M'}{M} \right)^2 
- \frac{1}{2 a^6 H^2 M^2} = 0
\;\; . \label{Meqn}
\end{equation}
We have converted from co-moving time $t$ to the number
of $e$-foldings from the beginning of inflation $n$:
\begin{equation}
n \equiv \ln \! \left[ \frac{a(t)}{a_i} \right]
\quad , \quad 
\frac{d}{dt} = H \frac{d}{dn}
\quad , \quad 
' \equiv \frac{d}{dn}
\;\; . \label{t_to_n}
\end{equation}
An evolution equation like (\ref{Meqn}) is preferable
since it avoids the need to take into account the 
oscillating phases of the mode functions that enter into 
the evolution equations (\ref{ueqns}-\ref{veqns}).
\\ [3pt]
{\it - Step 2: Decomposition into ``background" $\times$
``residual".} 
\\ [2pt]
The next step is to write our variable $M(t, k)$ in the form
of an appropriate background $M_0(t, k)$ times a residual
$h(n, k)$:
\begin{equation}
M(t,k) \equiv M_0(t, k) \times 
\exp \! \left[ - \frac12 h(n, k) \right]
\;\; . \label{M0_times_h}
\end{equation}
The background is chosen by requiring that it captures
the main effect. The residual is to be determined from 
the evolution equation it satisfies. 
\\ [3pt]
{\it - Step 3: The background choice.} 
\\ [2pt]
The background should incorporate the main effect 
by taking simultaneously into account: 
\\ [2pt]
{\it (i)} the relative success of the slow-roll 
approximation,
\\ [2pt]
{\it (ii)} the need to allow for time dependent 
$\epsilon(t)$,
\\ [2pt]
{\it (iii)} the constancy that the physical mode function 
$M(t, k)$ eventually achieves past first horizon crossing;
since $\epsilon(t)$ continues to evolve, so does $M_0(t, k)$
and its time dependence should be eliminated by a compensating 
time dependence in the residual $h(n, k)$ to obtain the 
required constancy of the full mode function $M(t, k)$.
\\ [2pt]
With these requirements in mind, we choose:
\begin{eqnarray}
\forall t < t_k &\!\!\! : &
M_0(t, k) \; {\rm from \; instantaneously \;
constant \; \epsilon \; solution}
\;\; , \label{M0<} \\ 
\forall t > t_k &\!\!\! : &
M_0(t, k) \; {\rm from \; constant \; \epsilon_k \; 
solution}
\;\; . \label{M0>}
\end{eqnarray}
This can be mathematically expressed as follows:
\begin{equation}
M_0(t, k) \, = \,
\theta(t_k - t) \; M_{\rm inst}(t, k) 
\, + \,
\theta(t - t_k) \; {\overline M}_{\rm inst}(t, k) 
\;\; , \label{M0}
\end{equation}
with the understanding that the instantaneously
constant $\epsilon$ solution is:
\begin{equation}
M_{\rm inst}(t,k) 
\, \equiv \,
\frac{z(t,k) \; \mathcal{H}( \nu(t), z(t,k) )}
{2 k \, a^2(t)} 
\quad , \quad
\mathcal{H}(\nu,z) \equiv 
\frac{\pi}{2} \Big\vert H^{(1)}_{\nu}(z)\Big\vert^2
\;\; , \label{Minst}
\end{equation}
with the usual definitions:
\begin{equation}
\nu(t) \equiv \frac12 + \frac{1}{1 - \epsilon(t)}  
\quad , \quad 
z(t,k) \equiv \frac{k}{[1 - \epsilon(t)] H(t) \, a(t)} 
\;\; . \label{inst}
\end{equation}
From first horizon crossing onwards, the backround 
is the constant $\epsilon$ solution for $\epsilon(t)
= \epsilon_k$:
\begin{equation}
{\overline M}_{\rm inst}(t,k) 
\, \equiv \,
\frac{{\overline z}(t,k) \; 
\mathcal{H}( \nu(t), {\overline z}(t,k) )}
{2 k \, {\overline a}^2(t)} 
\;\; . \label{Minst2}
\end{equation}
In terms of the number of $e$-foldings from first
horizon crossing $\; \Delta n \equiv n - n_k \,$
the geometrical parameters of (\ref{Minst2}) are:
\begin{equation}
\overline{a}(n) = a(n) = 
a_k e^{\Delta n} 
\quad , \quad 
\overline{H}(n) = 
H_k \, e^{-\epsilon_k \Delta n} 
\quad , \quad 
\overline{\epsilon}(n) = \epsilon_k 
\;\; . \label{fakegeom}
\end{equation}
\newpage
{\it - Step 4: The primordial tensor power spectrum.} 
\\ [2pt]
The late time limit of $M_0(t, k)$ is: 
\begin{equation}
\lim_{t \gg t_k} M_0(t, k) \, = \,
\lim_{t \gg t_k} 
\frac{{\bar z}(t, k)}{2 k \, {\bar a}^2(t)} \; 
{\cal H}( \nu(t), z(t, k) )
\, = \, 
\frac{H^2(n_k)}{2 k^3} \times C[\epsilon(n_k)]
\;\; . \label{M02}
\end{equation}
The physical object of interest is the tensor power spectrum
(\ref{Dh}) which in view of (\ref{M02}) now equals:
\begin{eqnarray}
\Delta^2_{h}(k) &\!\! = \!\!&
\frac{k^3}{2\pi^2} \times 32\pi G \times 2 \times 
M_0(t, k) \, e^{-\frac12 h(n, k)} \Big\vert_{t \gg t_k}
\;\; , \label{Dh2} \\
&\!\! = \!\!&
\frac{16 G H^2(n_k)}{\pi} \times C[\epsilon(n_k)]
\times e^{\tau[\epsilon](k)}
\;\; . \label{Dh3}
\end{eqnarray}
The non-local correction factor $\tau[\epsilon](k)$ 
to the tensor power spectrum is seen to be:
\begin{equation}
\tau[\epsilon](k) \, \equiv \,
\lim_{n \gg n_k} \left[ -\frac12 h(n, k) \right]
\;\; . \label{tau}
\end{equation}
\\ [-5pt]
{\it - Step 5: The residual evolution equation.} 
\\ [2pt]
In terms of the natural frequency of the system:
\footnote{When $\omega \sim 1$ we have about 
one oscillation per $e$-folding.}
\begin{equation}
\omega(n, k) \equiv
\frac{1}{a^3(t) \, H(t) \, M_0(t, k)}
\;\; , \label{omega}
\end{equation}
and upon substituting the generic relation 
(\ref{M0_times_h}), we can express (\ref{Meqn}) 
as follows:
\begin{equation}
h'' - \frac{\omega'}{\omega} h' + \omega^2 h
\, = \,
\frac14 h'^{\, 2} - \omega^2 \Big( e^h - 1 - h \Big)
\, + \, S_h
\;\; . \label{heqn}
\end{equation}
This is -- up to the non-linearities -- an equation
of a damped oscillator driven by the tensor source
$S_h$:
\begin{equation}
S_h \, \equiv \,
- 2 \left( \frac{\omega'}{\omega} \right)'
+ \left( \frac{\omega'}{\omega} \right)^{\!\! 2}
+ 2 \epsilon' - (3 - \epsilon)^2
+ \frac{4k^2}{a^2 H^2} - \omega^2
\;\; . \label{S_h}
\end{equation}
In \cite{Brooker:2015iya} we have been able to solve for the 
retarded Green's function $G_h(n ; m)$ of the 
linear differential operator $D_h$ that appears 
on the left hand side of (\ref{heqn}):
\begin{eqnarray}
D_h &\!\! \equiv \!\!&
\partial_n^2 - \frac{\omega'}{\omega} \partial_n 
+ \omega^2
\quad \Longrightarrow \quad
\label{D_h} \\
G_h(n ; m) &\!\! = \!\!&
\frac{\theta(n-m)}{\omega(m, k)} \;
\sin \! \left[ \int_0^n dn' \; \omega(n', k) \right]
\;\; . \label{G_h}
\end{eqnarray}
The above Green's function is exact and true 
for {\it any} choice of the background $M_0$.
As a result, we can perturbatively solve (\ref{heqn})
with initial value data $h(0, k) = h'(0, k) = 0$:
\begin{eqnarray}
& \mbox{} &
\hspace{-2.1cm}
h \, = \,
h_1 + h_2 + \dots
\;\; , \label{h} \\
h_1(n,k) &\!\! = \!\!& 
\int_0^{n} \!\! dm \, G_h(n;m) \, S_h(m,k) 
\;\; , \label{h1} \\
h_2(n,k) &\!\! = \!\!& 
\int_0^{n} \!\! dm \, G_h(n;m) 
\left\{ \frac14 \Big[ h_1'(m,k) \Big]^2 
- \frac12 \Big[ \omega(m,k) \, h_1(m,k)\Big]^2 \right\} 
\;\; . \qquad \label{h2} 
\end{eqnarray}
\\ [-7pt]
{\it - Step 6: The physical approximations.} 
\\ [2pt]
The non-linear terms in $h(n, k)$ can be safely ignored
because no model consistent with the scalar data gives
large values of either $h(n,k)$ or $h'(n,k)$. Thus, we 
shall only consider the first term (\ref{h1}) in the 
perturbative solution (\ref{h}).

Moreover, we should identify the measure of deviation 
from constant $\epsilon$ geometries and make a physical 
approximation that will enable us to achieve a reasonable 
analytic expression for $\tau[\epsilon](k)$ which is also 
accurate. This identification is interval dependent:
\begin{eqnarray}
\forall t < t_k &\!\! \Rightarrow \!\!&
M_0 \; {\rm depends \; on \; instantaneous} \; \epsilon(t)
\nonumber \\
&\!\! \Rightarrow \!\!&
{\rm measures \; of \; deviation \; are:} \;
\epsilon' \, , \, \epsilon'' \, , \, (\epsilon')^2
\;\; , \label{deviation1} \\
\forall t > t_k &\!\! \Rightarrow \!\!&
M_0 \; {\rm depends \; on \; constant} \; \epsilon_k
\nonumber \\
&\!\! \Rightarrow \!\!&
{\rm measure \; of \; deviation \; is:} \;
\Delta\epsilon(n) = \epsilon(n) - \epsilon(k)
\;\; . \label{deviation2}
\end{eqnarray}
The above deviation measures can be most easily seen by 
substituting the frequency (\ref{omega}) in the tensor 
source (\ref{S_h}) and noting that the resulting $S_h$ 
for $t < t_k$ contains terms proportional to $\epsilon'$,
$\epsilon''$, $(\epsilon')^2$ but not to $\Delta\epsilon(n)$,
while the reverse is true for $t > t_k$. The approximation 
consists of: 
\\ [2pt]
{\it (i)} First extracting the terms proportional to the 
measures of deviation (\ref{deviation1},\ref{deviation2}) 
from $S_h$.
\\ [2pt]
{\it (ii)} Then setting $\epsilon = 0$ throughout given 
that the range favoured by current data is $0 \leq \epsilon 
\leq 0.01$.

The approximated Green's function (\ref{G_h}) equals:
\begin{eqnarray}
\lim_{\epsilon = 0} G_h(n; m) &\!\! = \!\!&
\theta(n - m) \;
\frac12 e^{\Delta m} \Big( 1 + e^{2 \Delta m} \Big) \times
\nonumber \\
& \mbox{} &
\hspace{-1cm}
\times \, \sin \! \Bigg[ \! -2 \, \Big\{ e^{- \Delta l} 
- \arctan \! \Big( e^{- \Delta l} \Big) \Big\}
\Big\vert_m^n \Bigg]
\quad , \quad
\Delta m \equiv m - n_k
\;\; . \qquad \label{G_h2}
\end{eqnarray}
When concerned with the power spectrum, we must take the
late time limit of (\ref{G_h2}):
\begin{equation}
\lim_{n \gg 1}G_h(n; m) \, = \,
\frac12 e^{\Delta m} \Big( 1 + e^{2 \Delta m} \Big) \,
\sin \! \left[ 2 e^{- \Delta m} 
- 2 \arctan \! \Big( e^{- \Delta m} \Big) \right]
\;\; . \label{G_h3}
\end{equation}
Henceforth, for (\ref{G_h3}) we shall use the compact form:
\begin{equation}
G_h(x) \, = \,
\frac12 (x + x^3) \,
\sin \! \left[ \frac{2}{x} - 
2 \arctan \! \Big( \frac{1}{x} \Big) \right]
\quad , \quad
x \equiv e^{\Delta m} 
\;\; . \label{G_h4}
\end{equation} 

The approximated source is the sum of the contributions 
from the two time intervals and their interface:
\begin{eqnarray}
\forall t < t_k : &\! S_h \!\!\!& = 
-2 \, \Big\{ \epsilon''(n) \, {\cal E}_1(x) 
+ [ \epsilon'(n) ]^2 \, {\cal E}_2(x)
+ \epsilon'(n) \, {\cal E}_3(x) \Big\}
\;\; , \label{S_h<} \\
t = t_k : &\! S_h \!\!\!& = 
+2 \, \epsilon'(n_k) \, {\cal E}_1(1) \delta(n - n_k)
\;\; , \label{S_h=} \\
\forall t > t_k : &\! S_h \!\!\!& = 
+2 \, \Big\{ \Delta\epsilon(n) 
+ \frac{4 + 2 x^2}{1 + x^2} 
\! \int_{n_k}^{n} \! dm \; \Delta\epsilon(m) 
\Big\} \, \frac{2}{1 + x^2}
\;\; . \qquad \label{S_h>} 
\end{eqnarray}
The coefficient functions ${\cal E}_{1,2,3}$ are independent
of $\epsilon$ and are the following combinations of various 
derivatives of Hankel functions:
\begin{eqnarray}
\mathcal{E}_1(x) &\!\!\! = \!\!\!& 
-1 - \mathcal{A}_0(x) - \mathcal{B}_0(x) 
\;\; , \label{E1} \\
\mathcal{E}_2(x) &\!\!\! = \!\!\!& 
\frac12 - \mathcal{A}_0(x) - \mathcal{C}_0(x) 
- 2 \mathcal{D}_0(x) - \mathcal{E}_0(x) 
- \frac12 \Bigl[ 2 \!+\! \mathcal{A}_0(x) 
\!+\! \mathcal{B}_0(x)\Bigr]^2 
\;\; , \qquad \label{E2} \\
\mathcal{E}_3(x) &\!\!\! = \!\!\!& 
-1 + \mathcal{A}_0(x) \mathcal{B}_0(x) 
+ \mathcal{B}_0^2(x) 
+ 2 \mathcal{D}_0(x) + 2 \mathcal{E}_0(x) 
\;\; , \label{E3}
\end{eqnarray}
where we have defined:
\begin{eqnarray}
\mathcal{A}_0(x) &\!\! \equiv \!\!&
\lim_{\epsilon=0} \mathcal{A}(x)
\quad , \quad
\mathcal{A}(x) \, \equiv \,
\frac{\partial}{\partial \nu} \,
\ln \! \Big[ \mathcal{H} \Big( \nu,\frac{1}{x} \Big) 
\Big\vert^2 \;\; , \label{Ader} \\
\mathcal{B}_0(x) &\!\! \equiv \!\!& 
\lim_{\epsilon=0} \mathcal{B}(x)
\quad , \quad
\mathcal{B}(x) \, \equiv \,
- \frac{\partial}{\partial \ln(x)} \,
\ln \! \Big[ \mathcal{H} \Big( \nu,\frac{1}{x} \Big) 
\Big\vert^2 \;\; , \label{Bder} \\
\mathcal{C}_0(x) &\!\! \equiv \!\!& 
\lim_{\epsilon=0} \mathcal{C}(x)
\quad , \quad
\mathcal{C}(x) \, \equiv \,
\frac{\partial^2}{\partial \nu^2} \,
\ln \! \Big[ \mathcal{H} \Big( \nu,\frac{1}{x} \Big) 
\Big\vert^2 \;\; , \label{Cder} \\
\mathcal{D}_0(x) &\!\! \equiv \!\!& 
\lim_{\epsilon=0} \mathcal{D}(x)
\quad , \quad
\mathcal{D}(x) \, \equiv \,
- \frac{\partial^2}{\partial \ln(x) \partial \nu} 
\, \ln \! \Big[ \mathcal{H} \Big( \nu,\frac1{x} \Big) 
\Big\vert^2 \;\; , \label{Dder} \\
\mathcal{E}_0(x) &\!\! \equiv \!\!& 
\lim_{\epsilon=0} \mathcal{E}(x)
\quad , \quad
\mathcal{E}(x) \, \equiv \,
\frac{\partial^2}{\partial \ln(x)^{\, 2}} 
\, \ln \! \Big[ \mathcal{H} \Big (\nu,\frac1{x} \Big) 
\Big\vert^2 \;\; . \label{Eder} 
\end{eqnarray}
Unlike ${\cal B}_0$ and ${\cal E}_0$, the derivatives 
${\cal A}_0, {\cal C}_0, {\cal D}_0$ cannot be analytically 
expressed but have excellent approximations:
\begin{eqnarray}
\mathcal{A}_0(x) &\!\! \simeq \!\!&
\frac{1.5 x^2 + 1.8 x^4 - 1.5 x^6 + 0.63 x^8}{1 + x^2} 
\;\; , \label{Aappr} \\
\mathcal{B}_0(x) &\!\! = \!\!& 
\frac{-1 - 3 x^2}{1 + x^2} 
\;\; , \label{Bappr} \\
\mathcal{C}_0(x) &\!\! \simeq \!\!& 
\frac{x^2 + 6.1 x^4 - 3.7 x^6 + 1.6 x^8}{(1 + x^2)^2} 
\;\; , \label{Cappr} \\
\mathcal{D}_0(x) &\!\! \simeq \!\!& 
\frac{-3 x^2 - 6.8 x^4 + 5.5 x^6 - 2.6 x^8}{(1 + x^2)^2} 
\;\; . \label{Dappr} \\
\mathcal{E}_0(x) &\!\! = \!\!& 
\frac{4 x^2}{(1 + x^2)^2} 
\;\; . \label{Eappr} 
\end{eqnarray}
\\ [-7pt]
{\it - Step 7: The final answer for the tensor spectrum.} 
\\ [2pt]
In view of (\ref{tau},\ref{h},\ref{G_h4},\ref{S_h<}-\ref{S_h>}) 
the non-local correction factor equals:
\begin{eqnarray}
\tau[\epsilon](k) &\!\! = \!\!& 
\int_0^{n_k} \!\!\! dn \, \Bigg\{
\epsilon''(n) \, \mathcal{E}_1 (e^{\Delta n}) 
+ [ \epsilon'(n) ]^2 \, \mathcal{E}_2 (e^{\Delta n}) 
+ \epsilon'(n) \, \mathcal{E}_3(e^{\Delta n}) 
\Bigg\} \,  G(e^{\Delta n}) 
\nonumber \\
& \mbox{} &
\hspace{-0.8cm}
- \, \epsilon'(n_k) \, \mathcal{E}_1(1) \, G(1) 
\nonumber \\
& \mbox{} &
\hspace{-0.8cm}
- \!\! \int_{n_k}^{\infty}
\!\!\! dn \, \Bigg\{ \Delta\epsilon(n)
+ \frac{4 + 2e^{2 \Delta n}}{1 + e^{2 \Delta n}} 
\! \int_{n_k}^{n} \! dm \, \Delta\epsilon(m) \Bigg\} \,
\frac{2 \, G(e^{\Delta n})}{1 + e^{2 \Delta n}} 
\;\; , \label{taufinal}
\end{eqnarray}
and displays the tensor power spectrum dependence on the 
geometrical measures of deviation from constant $\epsilon$ 
backgrounds. 
\\ [7pt]
{\bf * The Scalar Spectrum:}
\\ [3pt]
{\it - Step 1: The optimal evolution variables.}
\\ [2pt]
From (\ref{veqns}) we can derive the following evolution 
equation in terms of the variable $N(t, k) \equiv 
\vert v(t, k) \vert^2$:
\begin{equation}
\frac{N''}{N} + 
\Big( 3 - \epsilon + \frac{\epsilon'}{\epsilon} \Big) 
\frac{N'}{N}
+ \frac{2 k^2}{a^2 H^2} 
- \frac12 \left( \frac{N'}{N} \right)^2 
- \frac{1}{2 a^6 H^2 \epsilon^2 N^2} = 0
\;\; . \label{Neqn}
\end{equation}
\\ [-7pt]
{\it - Step 2: Decomposition into ``background" $\times$
``residual".}
\\ [2pt]
We again write our variable $N(t, k)$ in the form of 
an appropriate background $N_0(t, k)$ times a residual
$g(n, k)$:
\begin{equation}
N(t,k) \equiv N_0(t, k) \times 
\exp \! \left[ - \frac12 g(n, k) \right]
\;\; . \label{N0_times_g}
\end{equation}
\\ [-7pt]
{\it - Step 3: The background choice.}
\\ [2pt]
The criteria for determining an optimal background
are identical to those employed for the tensor case.
The resulting choice for the background $N_0(t, k)$ 
is similar to (\ref{M0<}-\ref{M0>}):
\begin{eqnarray}
\forall t < t_k &\!\!\! : &
N_0(t, k) \; {\rm from \; instantaneously \;
constant \; \epsilon \; solution}
\;\; , \label{N0<} \\ 
\forall t > t_k &\!\!\! : &
N_0(t, k) \; {\rm from \; constant \; \epsilon_k \; 
solution}
\;\; , \label{N0>}
\end{eqnarray}
and can be mathematically expressed thusly:
\begin{equation}
N_0(t, k) \, = \,
\theta(t_k - t) \; N_{\rm inst}(t, k) 
\, + \,
\theta(t - t_k) \; {\overline N}_{\rm inst}(t, k) 
\;\; , \label{N0}
\end{equation}
with the understanding that, as before, the 
instantaneously constant $\epsilon$ solution is:
\begin{equation}
N_{\rm inst}(t,k) 
\, \equiv \,
\frac{z(t,k) \; \mathcal{H}( \nu(t), z(t,k) )}
{2 k \, \epsilon(t) \, a^2(t)} 
\quad , \quad
\mathcal{H}(\nu,z) \equiv 
\frac{\pi}{2} \Big\vert H^{(1)}_{\nu}(z)\Big\vert^2
\;\; , \label{Ninst}
\end{equation}
and the constant $\epsilon_k$ solution, appropriate 
after first horizon crossing, is:
\begin{equation}
{\overline N}_{\rm inst}(t,k) 
\, \equiv \,
\frac{{\overline z}(t,k) \; 
\mathcal{H}( \nu(t), {\overline z}(t,k) )}
{2 k \, \epsilon_k \, {\overline a}^2(t)} 
\;\; . \label{Ninst2}
\end{equation}
\\ [-7pt]
{\it - Step 4: The primordial tensor power spectrum.}
\\ [2pt]
Of physical interest is the late time limit $t \gg t_k$ 
of $N_0(t, k)$ is: 
\begin{eqnarray}
\lim_{t \gg t_k} N_0(t, k) &\!\! = \!\!&
\lim_{t \gg t_k} 
\frac{{\bar z}(t, k)}{2 k \, \epsilon_k \, {\bar a}^2(t)} \; 
{\cal H}( \nu(t), z(t, k) )
\nonumber \\
&\!\! = \!\!&
\frac{H^2(n_k)}{2 k^3} \, \epsilon_k \times C[\epsilon(n_k)]
\;\; , \label{N02}
\end{eqnarray}
from which the scalar power spectrum (\ref{Dh}) is
obtained:
\begin{eqnarray}
\Delta^2_{\mathcal{R}}(k) &\!\! = \!\!&
\frac{k^3}{2\pi^2} \times 4\pi G \times 2 \times 
N_0(t, k) \, e^{-\frac12 g(n, k)} \Big\vert_{t \gg t_k}
\;\; , \label{DR2} \\
&\!\! = \!\!&
\frac{16 G H^2(n_k)}{\pi \, \epsilon_k} \times C[\epsilon(n_k)]
\times e^{\sigma[\epsilon](k)}
\;\; . \label{DR3}
\end{eqnarray}
Therefore, the non-local correction factor $\tau[\epsilon](k)$ 
to the scalar power spectrum is:
\begin{equation}
\sigma[\epsilon](k) \, \equiv \,
\lim_{n \gg n_k} \left[ -\frac12 g(n, k) \right]
\;\; . \label{sigma}
\end{equation}
\\ [-5pt]
{\it - Step 5: The residual evolution equation.}
\\ [2pt]
In terms of the frequency of the system:
\begin{equation}
\Omega(n, k) \equiv
\frac{1}{a^3(t) \, H(t) \, \epsilon(t) \, N_0(t, k)}
\;\; , \label{Omega}
\end{equation}
the evolution equation (\ref{Neqn}) becomes:
\begin{equation}
g'' - \frac{\Omega'}{\Omega} g' + \Omega^2 g
\, = \,
\frac14 g'^{\, 2} - \Omega^2 \Big( e^g - 1 - g \Big)
+ S_g
\;\; . \label{geqn}
\end{equation}
As expected, we are again led to an equation describing
a damped oscillator -- with small non-linearities --
driven by the scalar source $S_g$:
\begin{equation}
S_g \, \equiv \,
- 2 \left( \frac{\Omega'}{\Omega} \right)'
+ \left( \frac{\Omega'}{\Omega} \right)^{\!\! 2}
- 2 \left( \frac{\epsilon'}{\epsilon} \right)^{\!\! 2} 
+ 2 \epsilon' 
- \Big( 3 - \epsilon + \frac{\epsilon'}{\epsilon} \Big)^{\! 2}
+ \frac{4k^2}{a^2 H^2} - \Omega^2
\;\; . \qquad \label{S_g}
\end{equation}
The solution for the retarded Green's function 
$G_g(n ; m)$ of the linear differential operator 
$D_g$ is similar:
\begin{eqnarray}
D_g &\!\! \equiv \!\!&
\partial_n^2 - \frac{\Omega'}{\Omega} \partial_n 
+ \Omega^2 
\quad \Longrightarrow \quad
\label{D_g} \\
G_g(n ; m) &\!\! = \!\!&
\frac{\theta(n-m)}{\Omega(m, k)} \;
\sin \! \left[ \int_0^n dn' \; \Omega(n', k) \right]
\;\; , \label{G_g}
\end{eqnarray}
and is valid for {\it any} expansion history \cite{Brooker:2016xkx}.
The perturbative solution to (\ref{geqn}) with initial
value data $g(0, k) = g'(0, k) = 0$ is:
\begin{eqnarray}
& \mbox{} &
\hspace{-2.1cm}
g \, = \,
g_1 + g_2 + \dots
\;\; , \label{g} \\
g_1(n,k) &\!\! = \!\!& 
\int_0^{n} \!\! dm \, G_g(n;m) \, S_g(m,k) 
\;\; , \label{g1} \\
g_2(n,k) &\!\! = \!\!& 
\int_0^{n} \!\! dm \, G_g(n;m) 
\left\{ \frac14 \Big[ g_1'(m,k) \Big]^2 
- \frac12 \Big[ \omega(m,k) \, g_1(m,k)\Big]^2 \right\} 
\;\; . \qquad \label{g2} 
\end{eqnarray}
\\ [-7pt]
{\it - Step 6: Relations between the tensor and the scalar case.}
\\ [2pt]
The scalar $\Omega$ and tensor $\omega$ frequencies are 
different but simply related:
\begin{equation}
\Omega(n, k) \, = \,
\theta(n_k - n) \, \omega(n, k) 
+ \theta(n - n_k) \, \omega(n, k) \, 
\frac{\epsilon_k}{\epsilon(n)}
\;\; . \label{omegas} 
\end{equation}
This implies the following relation between the corresponding
sources:
\begin{eqnarray}
\forall t < t_k : &\! S_g \!\!\!& = 
S_h -2 \left[ \Big( \frac{\epsilon'}{\epsilon} \Big)'
+ \frac12 \Big( \frac{\epsilon'}{\epsilon} \Big)^{\! 2} 
+ (3 - \epsilon) \frac{\epsilon'}{\epsilon} \right]
\;\; , \label{S_g<} \\
t = t_k : &\! S_g \!\!\!& =
S_h +2 \, \frac{\epsilon'}{\epsilon} \delta(n - n_k)
\;\; , \label{S_g=} \\
\forall t > t_k : &\! S_g \!\!\!& =
S_h -2 \left[ \Big( 3 - \epsilon 
+ \frac{\omega'}{\omega} \Big) \frac{\epsilon'}{\epsilon} 
+ \omega^2 \Big( \frac{\epsilon_k}{\epsilon} \Big)^{\! 2}
\right]
\;\; . \qquad \label{S_g>} 
\end{eqnarray}
\\ [-7pt]
{\it - Step 7: The physical approximations.}
\\ [2pt]
The two approximations that enable us to obtain
a simple analytic approximation for the primordial
power spectra are: 
\\ [2pt]
{\it (i)} the smallness of $\epsilon$ which led 
us to the forms (\ref{S_h},\ref{S_g}) for the 
sources, and
\\ [2pt]
{\it (ii)} the smallness of the non-linear terms
in (\ref{heqn},\ref{geqn}) which simplifies the
solution and eventually leads to (\ref{taufinal},
\ref{sigmafinal}).

In the case of the scalar spectrum the presence of
inverse factors of $\epsilon$ in the scalar source
(\ref{S_g}) makes those terms dominant relative to
the remaining $S_h$ term.
\footnote{Since $\epsilon < 0.01$ we expect $S_g$
to be about 100 times stronger than $S_h$.}
Therefore, the measures of deviation from constant 
$\epsilon$ geometries for the scalar case are:
\begin{equation}
\forall t \;\;
{\rm the \; measures \; of \; deviation \; are:} \quad
\frac{\epsilon'}{\epsilon} \; , \
\Big( \frac{\epsilon'}{\epsilon} \Big)^2 \, , \,
\Big( \frac{\epsilon'}{\epsilon} \Big)'
\;\; , \label{deviations}
\end{equation}

Now the compact form of the Green's function to be 
used for computing the scalar power spectrum is the 
same with that used in the tensor power spectrum 
(\ref{G_h4}) because: 
\begin{equation}
\lim_{\epsilon=0} G_h(n; m)
\, = \,
\lim_{\epsilon=0} G_g(n; m)
\;\; . \label{G_h=G_g}
\end{equation}
The approximated source is the sum of the contributions 
from the two time intervals and their interface:
\begin{eqnarray}
\forall t < t_k : &\! S_g \!\!\!& = 
-2 \left[ \Big( \frac{\epsilon'}{\epsilon} \Big)'
+ \frac12 \Big( \frac{\epsilon'}{\epsilon} \Big)^{\! 2} 
+ (3 - \epsilon) \frac{\epsilon'}{\epsilon} \right]
\;\; , \label{S_g<2} \\
t = t_k : &\! S_g \!\!\!& =
+2 \, \frac{\epsilon'}{\epsilon} \delta(n - n_k)
\;\; , \label{S_g=2} \\
\forall t > t_k : &\! S_g \!\!\!& =
-2 \left[ \Big( 3 - \epsilon 
+ \frac{\omega'}{\omega} \Big) \frac{\epsilon'}{\epsilon} 
+ \omega^2 \Big( \frac{\epsilon_k}{\epsilon} \Big)^{\! 2}
\right]
\;\; . \qquad \label{S_g>2} 
\end{eqnarray}
\\ [-7pt]
{\it - Step 8: The final answer for the scalar spectrum.}
\\ [2pt]
Therefore -- in view of
(\ref{sigma},\ref{g1},\ref{G_h4},\ref{S_g<2}-\ref{S_g>2})
-- the non-local correction factor equals:
\begin{eqnarray}
\sigma[\epsilon](k) &\!\! \simeq \!\!& 
\int_0^{n_k} \!\!\! dn \, \Bigg\{
\partial_n^2 \ln[\epsilon(n)]
+ \frac12 \Big( \ln[\epsilon(n)] \Big)^2
+ 3 \, \partial_n \ln[\epsilon(n)]
\Bigg\} \,  G(e^{\Delta n}) 
\qquad \nonumber \\
& \mbox{} &
\hspace{-0.8cm}
- \, \Big( \partial_{n_k} \ln[\epsilon(n_k)] \Big) \, G(1) 
\nonumber \\
& \mbox{} &
\hspace{-0.8cm}
- \!\! \int_{n_k}^{\infty}
\!\!\! dn \, \Big( \partial_n \ln[\epsilon(n)] \Big) \, 
\frac{2 \, G(e^{\Delta n})}{1 + e^{2 \Delta n}} 
\;\; , \label{sigmafinal}
\end{eqnarray}
and displays the scalar power spectrum dependence on the 
geometrical measures of deviation from constant $\epsilon$ 
backgrounds.

\begin{figure}[H]
\includegraphics[width=6.0cm,height=4.0cm]{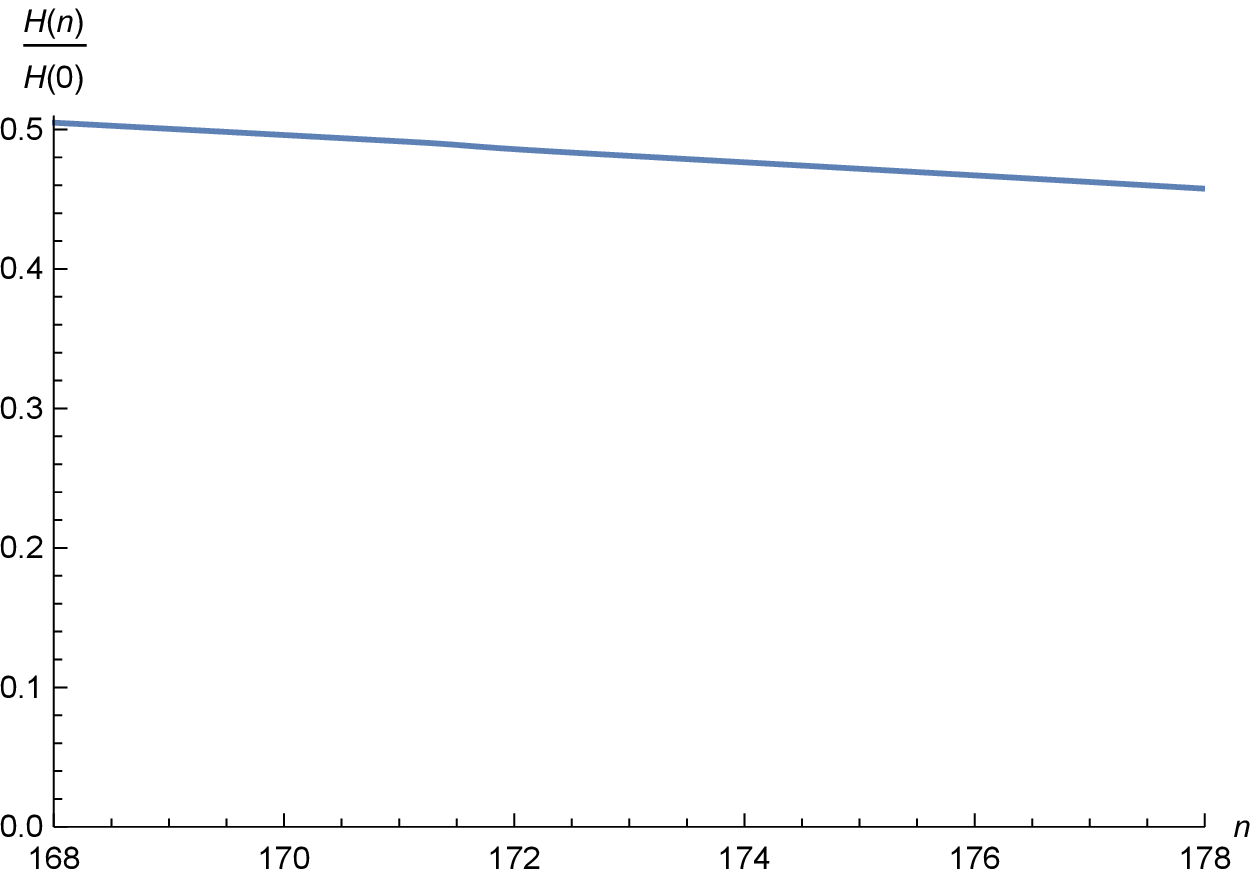}
\hspace{1cm}
\includegraphics[width=6.0cm,height=4.0cm]{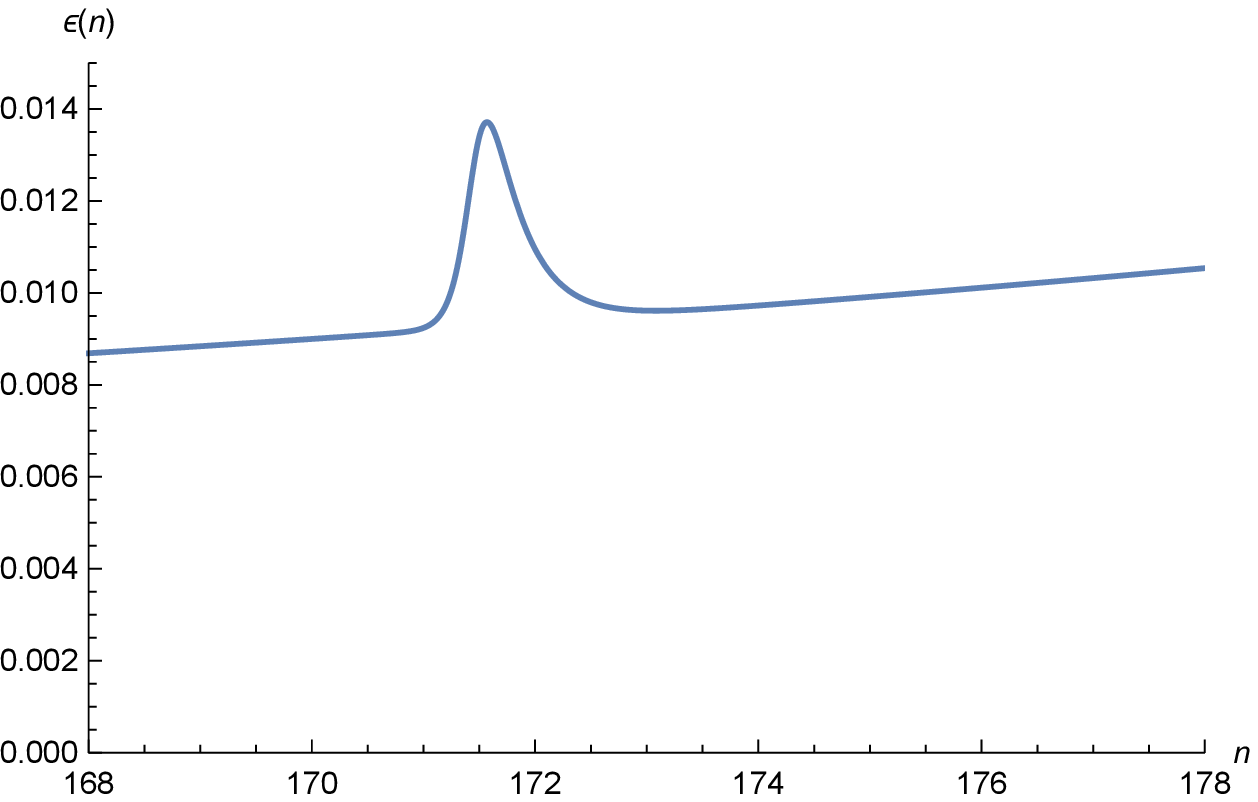}
\caption{\footnotesize The left hand figure shows the Hubble parameter 
and the right shows the first slow-roll parameter for a model with 
features. This model which was proposed \cite{Mortonson:2009qv,
Adams:2001vc} to explain the observed features in the scalar power 
spectrum at $\ell \approx 22$ and $\ell \approx 40$ which are visible 
in the data reported from both WMAP \cite{Covi:2006ci,Hamann:2007pa,
Hazra:2010ve} and PLANCK \cite{Hazra:2014goa,Hazra:2016fkm}. Note 
that the feature has little impact on $H(n)$ but it does lead to a 
distinct bump in $\epsilon(n)$.}
\label{StepGeometry}
\end{figure}

Finally, if desired the approximation (\ref{sigmafinal}) can be 
made even stronger by including the non-linear terms contained 
in (\ref{g2}). The process is straightforward although somewhat 
tedious. We first compute in the de Sitter limit the two nonlinear 
terms $\{g_1^2(n,k), (g'_1(n,k))^2\}$. They are found using the 
first order result (\ref{g1}) which is given before horizon 
crossing by:
\begin{eqnarray}
g_1(n<n_k,k)=\frac12\int_0^n \,dm\, S_g(m)e^{\Delta m}\left(1+e^{2\Delta m}\right)\nonumber\\
\sin\left[2\left\{e^{-\Delta m}-\tan^{-1}\left(e^{-\Delta m}\right)-e^{-\Delta n}+\tan^{-1}\left(e^{-\Delta n}\right)\right\}\right]\label{g1i}\\
g'_1(n<n_k,k)=\int_0^n \;dm\; S_g(m)e^{\Delta m}\frac{1+e^{2\Delta m}}{1+e^{2\Delta n}}\nonumber\\
\cos\left[2\left\{e^{-\Delta m}-\arctan\left(e^{-\Delta m}\right)-e^{-\Delta n}+\arctan\left(e^{-\Delta n}\right)\right\}\right]\label{g'1i}
\end{eqnarray}
where as before $\Delta n=n-n_k$. Taking the square of these two terms and inserting them into equation (\ref{g2}) yields the first nonlinear correction terms for $g(n,k)$. These correction terms are then to be viewed as source terms for $\sigma[\epsilon](k)$ and can be included in the integrand on the first line of equation (\ref{sigmafinal}).
\\ [7pt]
{\bf * The Power Spectrum Results:}
\\ [3pt]
It makes sense to apply the above results to the well 
established data of the primordial scalar power spectrum
reported from WMAP \cite{Covi:2006ci,Hamann:2007pa,Hazra:2010ve} 
and PLANCK \cite{Hazra:2014goa,Hazra:2016fkm}. As mentioned
earlier, the data shows features at $\ell \approx 22$ and
$\ell \approx 40$ of $3\sigma$ statistical significance.
If taken seriously \cite{Mortonson:2009qv}, these could be 
explained by a model \cite{Adams:2001vc} with the particular 
forms of the Hubble parameter $H(n)$ and first slow-roll 
parameter $\epsilon(n)$ shown in Figure~\ref{StepGeometry}.

\begin{figure}[H]
\includegraphics[width=6.0cm,height=4cm]{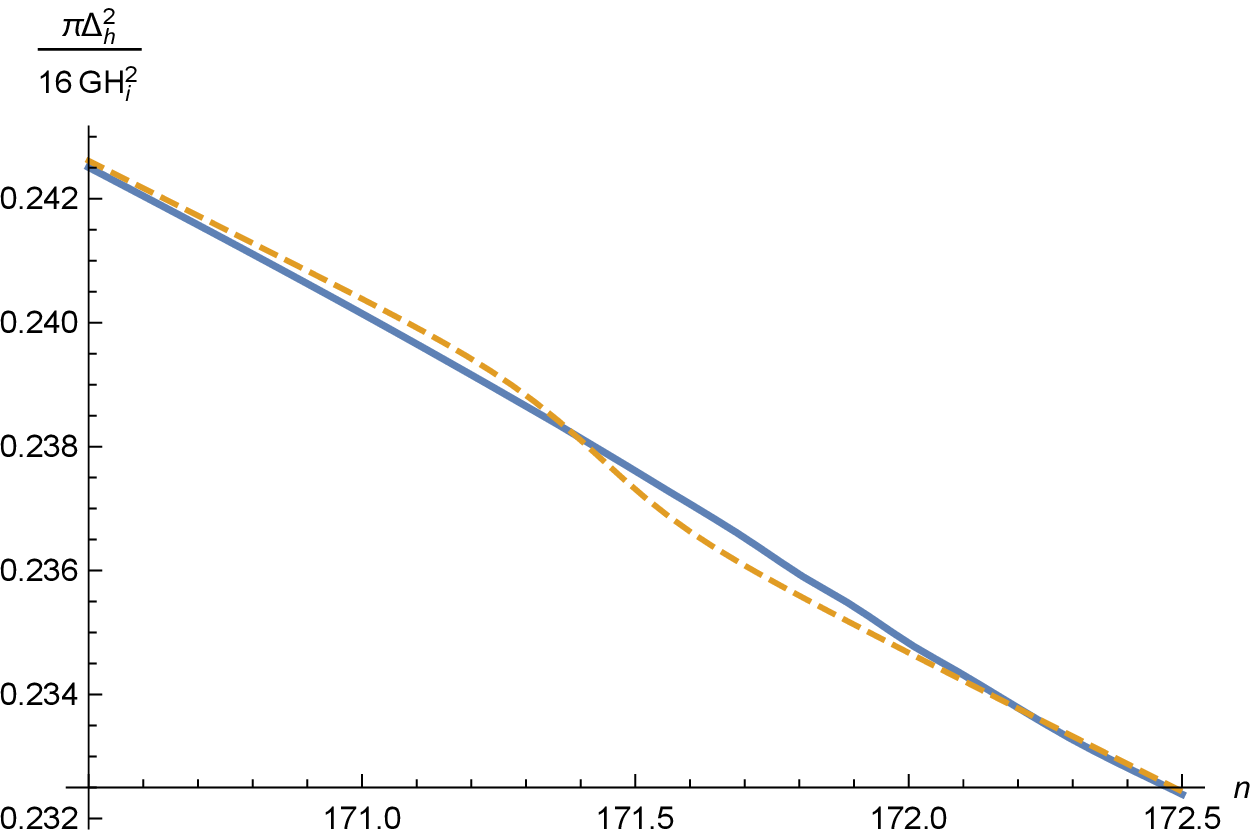}
\hspace{1cm}
\includegraphics[width=6.0cm,height=4cm]{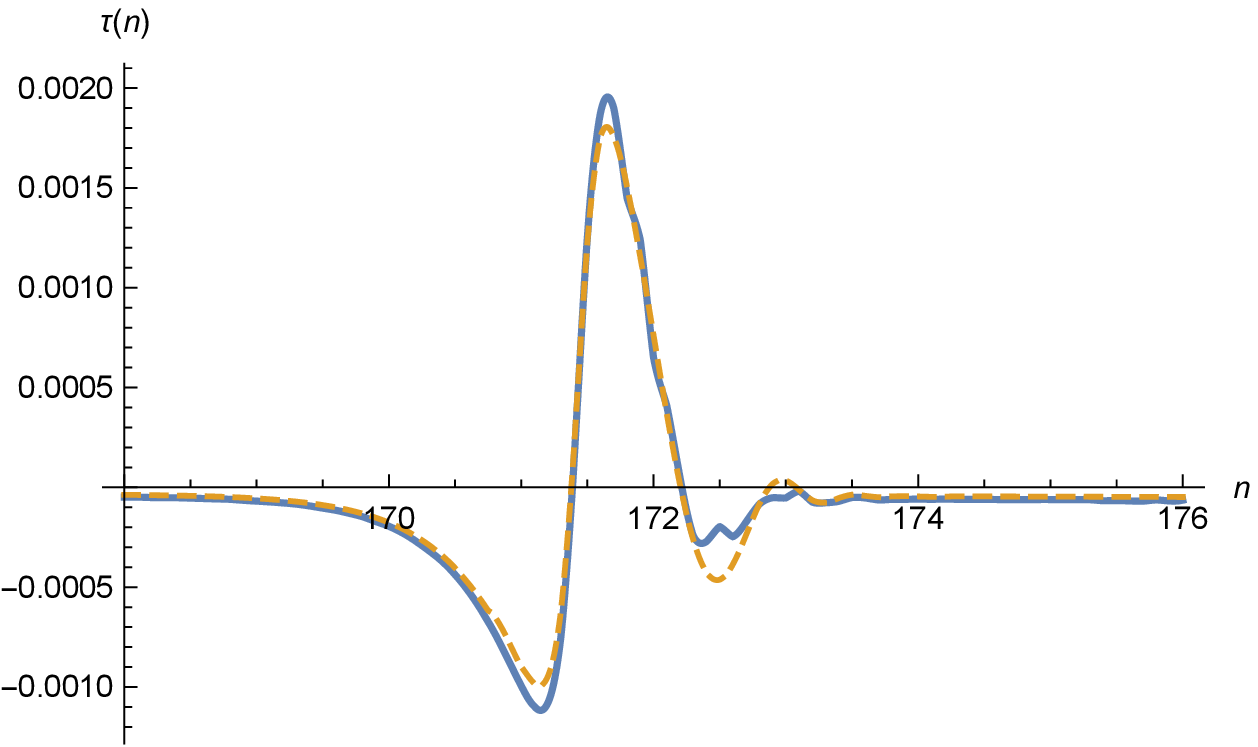}
\caption{\footnotesize These graphs show the tensor power 
spectrum for the model with a feature. The left hand figure 
compares the exact result {\it (solid blue)} with the local 
slow-roll approximation {\it (yellow dashed)}. The {\it solid 
blue} line on the right hand graph shows the logarithm of the 
ratio of the exact tensor power spectrum to its local slow-roll 
approximation. The {\it yellow dashed} line gives the non-local 
corrections of $\,\tau[\epsilon](k)$. The agreement is again 
excellent.}
\label{StepResultsTensor}
\end{figure}

The relevant results for the tensor power spectrum are best 
displayed in Figure~\ref{StepResultsTensor}. The agreement
between the numerically obtained exact result and our 
approximation (\ref{taufinal}) is almost perfect while the 
same is not true for the local slow-roll approximation. One
hopes that in the not too far future the tensor power spectrum
will be observed.

\begin{figure}[H]
\includegraphics[width=6.0cm,height=4.8cm]{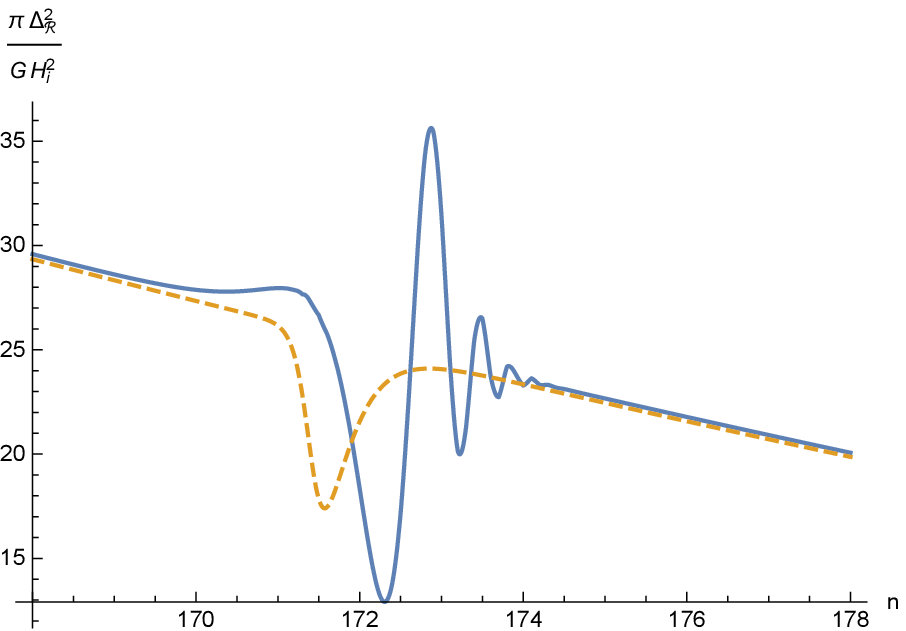}
\hspace{1cm}
\includegraphics[width=6.0cm,height=4.8cm]{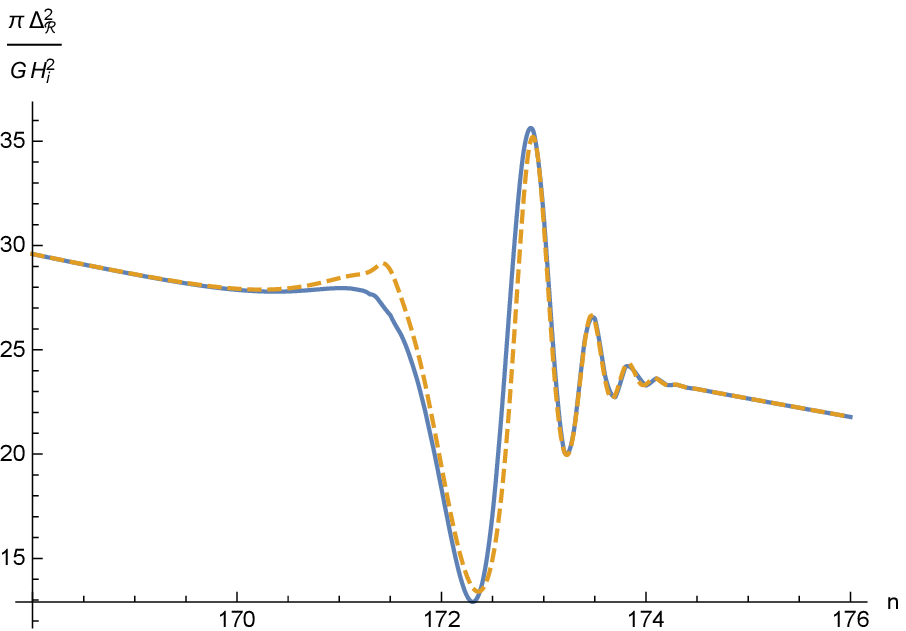}
\caption{\footnotesize These graphs show the results for the 
model of Figure~\ref{StepGeometry}. The left hand figure compares 
the exact scalar power spectrum {\it (solid blue)} with its local 
slow-roll approximation {\it (yellow dashed)}. The right hand 
figure compares the exact result {\it (solid blue)} with the much 
better approximation {\it (yellow dashed)} obtained from the full 
form (\ref{DR3}), with our analytic approximation (\ref{sigmafinal}) 
for $\sigma[\epsilon](k)]$. The local slow-roll approximation 
does not give a very accurate fit even to the main feature in 
the range $\, 171 < n < 172.5 \,$, and it completely misses the 
secondary oscillations visible in the range $\, 172.5 < n < 174$. 
There is also a small, systematic offset before and after the 
features. The non-local contributions of (\ref{sigmafinal}) are 
essential for correctly reproducing the actual power spectrum.}
\label{StepResultsScalar}
\end{figure}

\begin{figure}[H]
\begin{center}
\includegraphics[width=6.0cm,height=4.8cm]{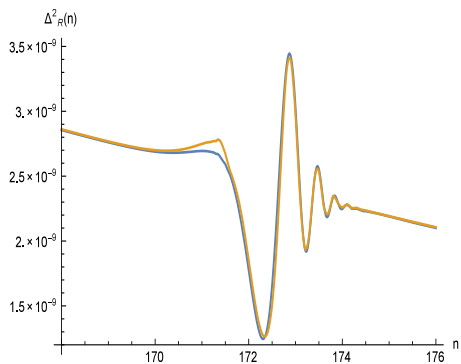}
\end{center}
\caption{\footnotesize This graph highlights the improvement made 
by adding the first nonlinear corrections to (\ref{sigmafinal}). 
In This figure the horizontal axis counts the number of e-foldings 
until the end of inflation. The most dramatic improvement is near 
$n=172$.} 
\label{nonlin}
\end{figure}

On the other hand, the scalar power spectrum signal is much 
stronger and has already been seen. In Figure~\ref{StepResultsScalar} 
we present the results for the model of \cite{Mortonson:2009qv,
Adams:2001vc}. There we can see the numerically obtained exact 
scalar power spectrum versus the local slow roll approximation 
(\ref{Dlocal}), and versus our analytic approximation 
(\ref{sigmafinal}). 
\noindent The analytic approximation can be made even better by 
including the first nonlinear corrections to $g(n,k)$ as shown in 
Figure~\ref{nonlin}. 
\noindent In view of these results itt becomes evident that:
\\ [3pt]
$\bullet \;$ The local slow-roll approximation 
(\ref{Dlocal}) is not accurate in reproducing the power
spectrum since it only reproduces the main oscillation 
with quite smaller amplitude, misses all the oscillations 
that follow and exhibits an offset throughout.
\\ [3pt]
$\bullet \;$ Our analytic non-local approximation 
(\ref{sigmafinal}) accurately follows the exact power spectrum 
over the whole range including the ``ringing'' after the main 
oscillation.

The secondary oscillations in the scalar power spectrum are
attributed to the presence of the feature which implies that 
$\epsilon'(n) \; \& \; \epsilon''(n) \neq 0$. Hence, there 
is deviation from the slow-roll approximation and this is 
imprinted in the source. Recall that the equation obeyed by 
the residual $g(n, k)$ is that of a damped driven oscillator:
\begin{equation}
g'' - \frac{\Omega'}{\Omega} g' + \Omega^2 g 
\, = \, S_g
\;\; \label{ringing}
\end{equation}
The restoring force in the oscillator (\ref{ringing}) is
exponentially proportional to $(-\Delta n) = - (n - n_k)$; 
thus:
\\ [2pt]
{\it (i)} For $t < t_k$ it is exponentially big and overwhelms
any effect from $S_g$.
\\ [2pt]
{\it (ii)} For $t > t_k$ it is exponentially small, as is 
$S_g$, and we have no ringing.
\\ [2pt]
{\it (iii)} For $t \approx t_k$ all forces are of $O(1)$ and 
we have the ringing from a damped driven oscillator.

In one sentence, the local slow-roll approximation cannot 
capture the effect of features since its only support is 
at $t_k$ while a feature leads to transient tails which 
need non-local support and our analytic approximation 
provides just that.

\section{From the CMBR to the Geometry}

\begin{figure}[ht]
\begin{center}
\includegraphics[width=6.0cm,height=4.8cm]{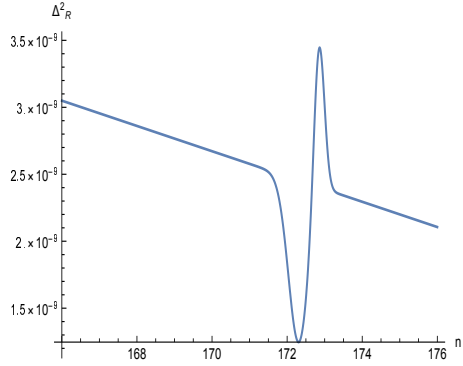}
\end{center}
\caption{\footnotesize The spectrum $\Delta_{\mathcal{R}}^2$ whose 
geometry we will reconstruct.}
\label{D2bump}
\end{figure}

We shall now consider the inverse problem of reconstructing
the geometry $H(n)$ and $\epsilon(n)$ from the power spectra
data $\Delta^2_{h}(k)$ and $\Delta^2_{\mathcal{R}}(k)$.
Before doing so there are some remarks that must be 
highlighted: 
\\ [3pt]
$\bullet \;$ It is an experimental fact that while the scalar 
power spectrum $\Delta^2_{\mathcal{R}}(k)$ is very well 
measured, the tensor power spectrum $\Delta^2_{h}(k)$ has yet 
to be resolved and, even when detected, it will be years before
much precision is attained. Therefore, reconstruction 
should be based on $\Delta^2_{\mathcal{R}}(k)$.
\footnote{The tensor spectrum $\Delta^2_{h}(k)$ is used only 
to fix the integration constant which gives the scale of 
inflation.}
\\ [3pt]
$\bullet \;$  The observed smallness of $\epsilon(n)$ and 
its assumed smoothness -- up to small transients responsible 
for $\epsilon'(n) \neq 0$ -- motivates a hierarchy between 
$H$, $\epsilon$ and $\frac{\epsilon'}{\epsilon}$ based 
on calculus:
\begin{equation}
H(n) = H_i \, \exp \Bigl[-\!\! \int_0^n \!\! dm \; 
\epsilon(m)\Bigr] 
\quad , \quad
\epsilon(n) = \epsilon_i \exp \Bigl[ \int_0^n \!\! dm \; 
\frac{\epsilon'(m)}{\epsilon(m)} \Bigr] 
\;\; . \label{sequence} 
\end{equation}
Hence $H(n)$ is insensitive to small errors in $\epsilon(n)$, 
and $\epsilon(n)$ is insensitive to small errors in 
$\partial_n \ln[\epsilon(n)]$.
\\[3pt]
We will demonstrate the reconstruction algorithm by applying it 
to a new toy model where the scalar spectrum is made to mimic 
the present data by having two large features but being otherwise 
flat.The functional form of the scalar spectrum which we consider 
is:
\begin{eqnarray}
\Delta_{\mathcal{R}}^2(N_k) = & 19.08 \times 10^{-9} - 9.65 \times 
10^{-11} n_k -1.21 \times 10^{-9} \, e^{-7(172.296 - n_k)^2} 
\nonumber \\
& + 1.18 \times 10^{-9} \, e^{-26(172.85 - n_k)^2}
\;\;. \label{dmock}
\end{eqnarray}
where $n_k$ is the number of e-foldings from the start of inflation,
and inflation ends at $n_e = 225.626$. 
A graph of this spectrum is shown in Figure~\ref{D2bump}. We will 
imagine that the tensor amplitude is $\Delta_h^2(n_k = 165.626) = 
3.1 \times 10^{-11}$ so that $H_i$ has the nominal value $2.8 \times 
10^{-5}/\sqrt{8\pi G}$ at a time 60 e-foldings before the end of 
inflation. We stress that the exact time (or wave number) at which 
we fix $H_i$ is inconsequential. In the event of a positive detection 
of primordial B modes we will use whichever wave number has the most 
well determined value for the tensor amplitude.
\\ [7pt]
{\bf * Reconstructing $H(n)$ and $\epsilon(n)$:}
\\ [3pt]
{\it - Step 1: The optimal variables.}
\\ [2pt]
It will be convenient to use dimensionless
variables for our purposes by dividing out by the
inflationary scale $H_i \equiv H(n=0)$
\begin{equation}
h(n) \, \equiv \,
\frac{H(n)}{H_i} 
\qquad , \qquad 
\delta(n_k) \, \equiv \,
\frac{\pi \Delta^2_{\mathcal{R}}(k)}{G H_i^2} 
\;\; , \label{recon}
\end{equation}
where we recall that $n_k$ is the number of $e$-foldings 
from the beginning of inflation to first horizon crossing 
for the wave number $k$.
\\ [3pt]
{\it - Step 2: The reconstruction formula.}
\\ [2pt]
Starting from the exact expression (\ref{DR3}) for
$\Delta^2_{\mathcal{R}}(k)$ we note that:
\\ [2pt]
{\it (i)} Because the observed first slow-roll 
parameter $\epsilon$ is very small, the local slow-roll 
correction factor $C[\epsilon_k]$ can be very safely 
ignored (see Figure~\ref{C(e)}). 
\\ [2pt]
{\it (ii)} Because the approximation (\ref{sigmafinal}) 
to the non-local correction exponent is superb, it can
be very safely used for our purposes (see 
Figure~\ref{StepResultsScalar}).
\\ [2pt]
Hence, expression (\ref{DR3}) reduces to the following
equation:
\begin{equation}
\delta(n) \, \simeq \,
\frac{h^2(n)}{\epsilon(n)} \times 
\exp \left[ \sum_{i=1}^5
{\rm exp}_i(n) \right] 
\;\; , \label{reconeqn1}
\end{equation}
which shall form the basis of our reconstruction technique.
The five exponents defined in (\ref{reconeqn1}) are the
various terms contained in (\ref{sigmafinal}):
\begin{eqnarray}
{\rm exp}_1(n) &\!\! = \!\!& 
-\partial_{n} \ln[\epsilon(n)] \times G(1) 
\;\; , \label{exp1} \\
{\rm exp}_2(n) &\!\! = \!\!& 
\int_0^{n} \!\! dm \; \partial_{m}^2 \ln[\epsilon(m)] 
\times G(e^{m-n}) 
\;\; , \label{exp2} \\
{\rm exp}_3(n) &\!\! = \!\!& 
\frac12 \! \int_0^{n} \!\! dm \; 
\Big[ \partial_m \ln[\epsilon(m)] \Big]^2 
\times G( e^{m-n}) 
\;\; , \label{exp3} \\
{\rm exp}_4(n) &\!\! = \!\!& 
3\! \int_0^{n} \!\! dm \; \partial_{m} \ln[\epsilon(m)] 
\times G(e^{m-n}) 
\;\; , \label{exp4} \\
{\rm exp}_5(n) &\!\! = \!\!& 
2\! \int_{n}^{\infty} \!\! dm \; \partial_m \ln[\epsilon(m)] 
\times \frac{G(e^{m-n})}{1 + e^{2(m-n)}} 
\;\; . \label{exp5}
\end{eqnarray}
\\ [-7pt]
{\it - Step 3: Reconstructing $H(n)$.}
\\ [2pt]
The reconstruction of the Hubble parameter is quite
simple. We ignore all the exponents in the
relevant equation (\ref{reconeqn1}) and only keep the
leading slow-roll terms:
\begin{equation}
\delta(n) \, \simeq \,
\frac{h^2(n)}{\epsilon(n)} 
\quad \Longrightarrow \quad
h^2(n) \, \simeq \,
\frac{1}{1 \!+\! \int_{0}^{n} \! dm \;\frac{2}{\delta(m)}}
\;\; . \label{slowrollH}
\end{equation}
Applying (\ref{slowrollH}) yields results which are excellent for 
fitting the flat parts of the spectrum. This interpolation is 
limited however since it does not contain any of the nonlocal 
character of the scalar spectrum and hence will not reproduce the 
features in the spectrum (\ref{dmock}). We will construct an even 
better interpolation of $h(n)$ at the end when we integrate our 
reconstructed $\epsilon(n)$ according to equation (\ref{sequence}). 
\\ [3pt]
{\it - Step 3: Reconstructing $\epsilon(n)$.}
\\ [2pt]
The case of the reconstruction of $\epsilon(n)$ is much
more delicate. We start by noting that in general $\{ {\rm exp}_1(n)$, 
${\rm exp}_2(n)$ and ${\rm exp}_4(n) \}$ are much larger than 
$\{ {\rm exp}_3(n)$ and ${\rm exp}_5(n) \}$ for models with transient 
features like the ones under consideration. The reason for this is 
that any deviations from slow roll which make ${\rm exp}_3(n)$ large 
will necessarily make the other terms  ${\rm exp}_2(n)$ and 
${\rm exp}_4(n)$ even larger by about an order of magnitude 
\cite{Brooker:2017kjd}. Moreover, as can be seen in (\ref{exp5}) the 
term ${\rm exp}_5$(n) is supressed relative to the other terms by the 
factor $(1+e^{2(m-n)})^{-1}$ in the integrand. These important facts 
make the optimal distribution of the five exponents in (\ref{reconeqn1}) 
transparent; upon taking the logarithm we get:
\begin{eqnarray}
- \ln[\epsilon(n)] - {\rm \exp}_1(n) 
- {\rm \exp}_2(n) - {\rm \exp}_4(n) 
&\!\! \simeq \!\!&
\nonumber \\
& \mbox{} &
\hspace{-3.9cm}
- \ln[\delta(n)] + 2 \ln[h(n)] 
+ {\rm \exp}_3(n) + {\rm \exp}_5(n)
\;\; . \qquad \label{reconeqn2}
\end{eqnarray}
It should be apparent that we cannot solve (\ref{reconeqn2})
exactly and we must develop an approximation technique.
One such technique is an iterative procedure with the 
following reconstruction logic:
\\ [2pt]
$\bullet \;$ We define the source ${\cal S}_{\epsilon}(n)$
as:
\begin{equation}
{\cal S}_{\epsilon}(n)
\, \equiv \,
- \ln[\delta(n)] + 2 \ln[h(n)] 
+ {\rm \exp}_3(n) + {\rm \exp}_5(n)
\;\; . \label{reconsource}
\end{equation}
The first term in (\ref{reconsource}) is determined from 
the measured primordial spectrum which gives $\delta(n)$. 
\\ [2pt]
$\bullet \;$ As the lowest order source that can provide
a lowest order solution we can take:
\begin{equation}
{\cal S}_0(n)
\, \equiv \,
- \ln[\delta(n)] - \ln \left[ 
1 + \int_{0}^{n} \!\! dm \; \frac{2}{\delta(m)} \right] 
\;\; . \label{reconsource0}
\end{equation}
Here we have determined the second term in 
(\ref{reconsource}) from the slow-roll formula
(\ref{slowrollH}) which gives $h(n)$ in terms of 
$\delta(n)$; a decent approximation because the
Hubble parameter does not change much due to the
presence of the feature. We have also ignored the
two weak terms ${\rm exp}_3(n)$ and ${\rm exp}_5(n)$.
\\ [2pt]
$\bullet \;$ The resulting lowest order equation:
\begin{equation}
\Big[ 1 + G(1) \partial_n \Big] \ln[\epsilon(n)] 
- \int_0^{n} \!\! dm \; 
\Big[ \partial_m^2 + 3 \partial_m \Big] 
\ln[\epsilon(m)] \times G(e^{m-n}) 
\, \simeq \,
{\cal S}_0(n)
\;\; , \label{reconeqn3}
\end{equation}
is a linear non-local equation that can be solved by
the Green's function method:
\begin{equation}
\ln[\epsilon(n)] = \int_0^{\infty} \! dm \; 
\mathcal{G}(n \!-\! m) \times {\cal S}_0(m) 
\;\; , \label{greensol}
\end{equation}
where the Green's function $\mathcal{G}(n)$ is the
solution to (\ref{reconeqn3}) for a delta function 
source:
\footnote{The function $\mathcal{G}$ becomes symmetric
in its arguments because the function $G(e^{n - n_k})$
given by (\ref{G_h4}) is essentially zero for
the interval up to $N \sim 4$ $e$-foldings before first
horizon crossing.}
\begin{equation}
\Big[ 1 + G(1) \partial_n \Big] \mathcal{G}(n) 
- \int_{-N}^{n} \!\! dm \;
\Big[ \partial_m^2 + 3 \partial_m \Big] 
\mathcal{G}(m) \times G(e^{m-n}) 
\, = \,
\delta(n) 
\;\; . \label{greeneqn}
\end{equation} 

\begin{figure}[H]
\includegraphics[width=6.0cm,height=4.8cm]{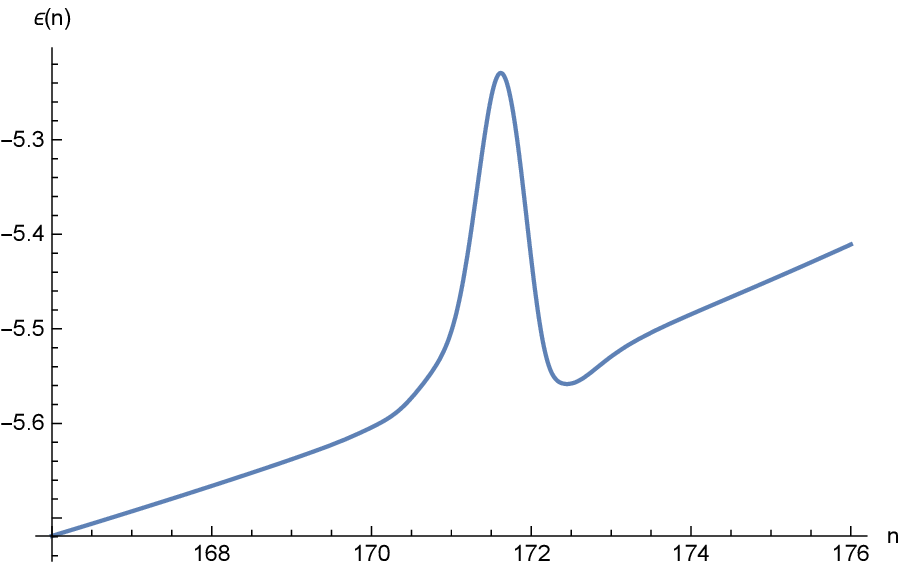}
\hspace{1cm}
\includegraphics[width=6.0cm,height=4.8cm]{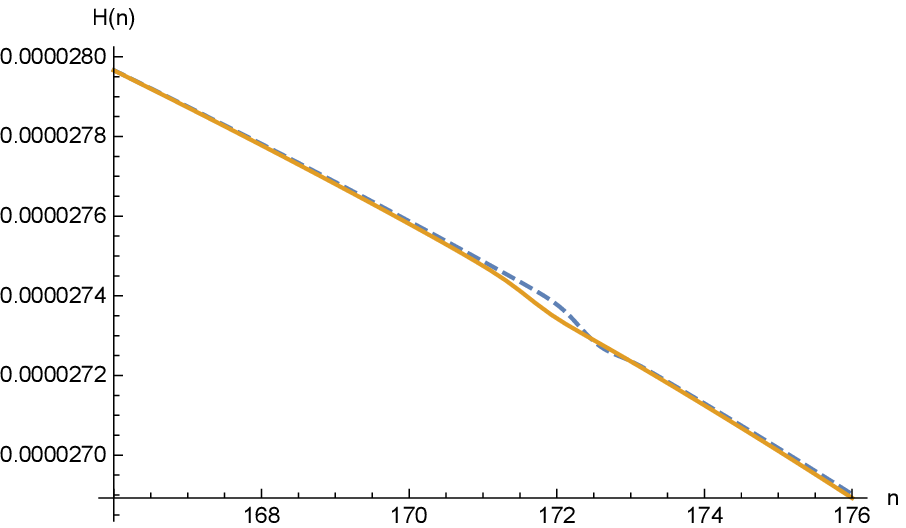}
\caption{\footnotesize The left hand graph shows the values of 
$\ln\left[\epsilon(n)\right]$ which we have reconstructed from 
(\ref{dmock}). On the right hand side are the two different 
interpolations of the Hubble parameter $H(n)$. The orange curve 
is the result of integrating $\epsilon (n)$ according to 
\ref{geometry} while the blue dashed curve comes from the 
leading slow roll terms in (\ref{slowrollH}). Note that the 
curves agree on the edges of the figure but disagree near the 
feature.}
\label{RecResults}
\end{figure}

The presence of the function $G(e^{n - n_k})$ 
in (\ref{greeneqn}) makes solving exactly for 
$\mathcal{G}(n)$ elusive. For the sake of simplicity 
it is easier to approach this problem after taking a 
Laplace transform because this turns our integro-differntial 
equation into an algebraic one. In the Laplace domain the 
Green's function equation we wish to solve is:
\begin{equation}
\Bigl[1 \!+\! G(1) s \!-\! (s \!+\! 3) s \!\times\! \mathcal{I}(s)\Bigr]
\widehat{\mathcal{G}}(s;m) = e^{-m s} \; , \label{Laplacegreen}
\end{equation}
where $\widehat{\mathcal{G}}(s;m)$ is the retarded Green's 
function we wish to find and we have defined, 
\begin{equation}
\mathcal{I}(s) \equiv \int_{0}^{\infty} \!\! d\ell \, e^{-s \ell} \!\times\!
G(e^{-\ell}) \; . \label{INT}
\end{equation}
We have show previously that a very good approximation for $\mathcal{I}(s)$ is given by:
\begin{eqnarray}
\mathcal{I}(s)=\mathcal{I}_0(s)+\mathcal{I}_1(s)\label{i0i1}\\
\mathcal{I}_0(s) = \frac{G(1)}{s} \Bigl[1 - e^{-0.8 s}\Bigr] \label{INT0}\\
\mathcal{I}_1(s) = \frac{0.154}{(s + 8.97)^2} \, \sin\Biggl[1.76 \Bigl(
1 \!-\! e^{-0.262 (s - 3.78)}\Bigr)\Biggr]\label{INT1}
\end{eqnarray}
for real values of $s$ \cite{Brooker:2017kjd}. Since finding the proper 
inverse transform is rather complicated, we simplify the process by 
expanding $\mathcal{I}(s)$ into its Taylor series. The solution for 
$\widehat{\mathcal{G}}$ is then expanded as a geometric series and 
the inverse transform is done term by term using the identity:
\begin{equation}
\mathcal{L}^{-1}\left[\frac{e^{-a\times s}}{(s+b)^q}\right] = 
\frac{e^{-b(a+n)}\Theta\left(a+n\right)\times\left(a+n\right)^{q-1}}{\Gamma(n)}
\end{equation}
The results for reconstructing the geometry of our mock spectrum 
are show in Figure~\ref{RecResults}.

\begin{figure}[H]
\begin{center}
\includegraphics[width=6.0cm,height=4.8cm]{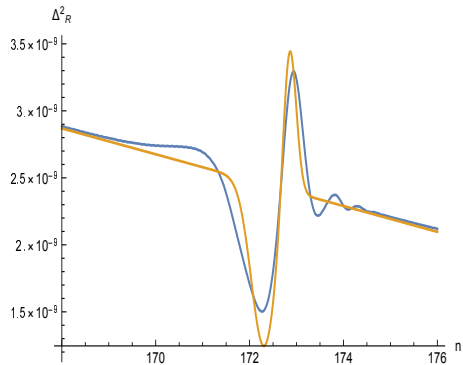}
\end{center}
\caption{\footnotesize The spectrum from the reconstructed geometry of 
Figure~\ref{RecResults} compared with the mock spectrum we started with.}
\label{SpecCheck}
\end{figure}

What we have found is that in order to produce a spectrum with exactly 
two features one would require two features in the geometry as opposed 
to the usual case where only one is considered. The first peak induces 
the ringing in the system just like we see in the step model however in 
this case the slight dip after the peak has the effect of canceling out 
the secondary peaks. We perform a check of our reconstruction by 
integrating $\epsilon_0(n)$ to obtain a new value of $H(n)$ and then 
inserting both into equation (\ref{DR3}) and comparing with our original 
model. This can be seen in Figure~\ref{SpecCheck} and the results speak 
for themselves.
We must stress that what we have done here is truely novel. We have not 
proposed any analytic model for either the potential or the geometry. We 
have explicitly constructed an excellent approximation to the geometry 
which would produce a model spectrum which mimics current data. 

\section{Epilogue}

In this paper we have presented a formalism that 
is applicable to geometries that include the presence 
of non-trivial features in their history. The formalism 
enables us to obtain analytic expressions for the 
primordial power spectra and also reconstruct the 
geometry given the power spectra as input.

The formalism determines the tree order power spectra 
by evolving the norm-squared mode functions. Even if 
considered purely as a numerical technique this is 
more efficient than evolving the mode functions because 
it avoids keeping track of the rapidly fluctuating phase, 
and because it converges about twice as fast. Moreover, 
the formalism applies not only to single-scalar inflation 
but also to any conformally related model, such as $f(R)$ 
inflation \cite{Brooker:2016oqa} and Higgs inflation, whose 
power spectra are numerically identical.

The methodology breaks down if the first slow-roll 
parameter $\epsilon$ becomes big, or if the non-linear
terms in the evolution equations become big. The power 
spectra results presented show that:
\\ [2pt]
{\it (i)} The slow-roll approximation breaks down when
the geometry contains non-trivial features.
\\ [2pt]
{\it (ii)} The non-local correction exponents 
$\tau[\epsilon]$ and $\sigma[\epsilon]$ are essential
in reaching quantitative accuracy.

In reconstructing the geometry from the power spectra 
data, our results indicate that further improvement
is needed for the accurately handling of derivatives
of the first slow-roll parameter $\epsilon$; as far
as the undifferentiated $\epsilon$ is concerned our
errors are accurate within $\pm 2.2\%$ and for the
Hubble parameter they never exceed $0.04\%$. These
results seem to improve on other techniques 
\cite{Kadota:2005hv,Barrow:2016wiy,Mastache:2016ahe}.

Another application concerns improving the classic
consistency relation \cite{Polarski:1995zn,
GarciaBellido:1995fz,Sasaki:1995aw} for comparing the
tensor power spectrum (when it is finally resolved) with
its well-measured scalar counterpart to test single-scalar
inflation. We have proposed a modification 
\cite{Brooker:2016imi} which:
\\ [2pt]
{\it (i)} Avoids the need to take a derivative of
$\Delta^2_h(k)$ that would degrade the accuracy of the 
poorly resolved initial detections.
\\ [2pt]
{\it (ii)} Integrates the high quality data we already
possess for $\Delta^2_{\mathcal R}(k)$.
\\ [2pt]
{\it (iii)} Can be used to cross-correlate scalar features
(e.g., Fig.~\ref{StepResultsScalar}) with the tensor 
features (e.g., Fig.~\ref{StepResultsTensor}), in the far 
future, when both spectra are well resolved.

A particularly exciting application of our 
formalism is to exploit the control it gives over how 
the mode functions depend upon $\epsilon(n)$ to design 
a new statistic to cross-correlate features in the 
power spectrum with non-Gaussianity. This has already 
been proposed in the context of models with variable 
speed of sound \cite{Achucarro:2012fd,Torrado:2016sls}, 
and developed numerically \cite{Hazra:2012yn}, but it can 
now be done analytically for simple scalar potential models. 
The idea is that non-Gaussianity measures self-interaction, 
which is what a step in the potential provides. There may 
be an observable effect which is not resolvable by generic 
statistics but could be detected by a precision search.

Finally, we mention using the formalism to motivate
better phenomenological models \cite{Tsamis:2014hra} of
the late time regime of $\Lambda$-driven inflation
\cite{Tsamis:2011ep,Tsamis:1996qq}. The fundamental assumption 
is that quantum gravitational back-reaction grows like the 
coincidence limit of the tensor propagator, which can be 
expressed as an integral of $M(t,k)$ \cite{Romania:2012av}. 
Inferring how this quantity depends on a general geometry 
defines the model.

\vskip 1cm 

\centerline{\bf Acknowledgements}

This work was partially supported by the European 
Union's Horizon 2020 Programme under grant agreement 
669288-SM-GRAV-ERC-2014-ADG; by NSF grant PHY-1506513; 
and by the Institute for Fundamental Theory at the 
University of Florida.

\end{document}